\documentclass{article}

\usepackage{PRIMEarxiv}

\usepackage[utf8]{inputenc} 
\usepackage[T1]{fontenc}    
\usepackage{hyperref}       
\usepackage{url}            
\usepackage{booktabs}       
\usepackage{amsfonts}       
\usepackage{nicefrac}       
\usepackage{microtype}      
\usepackage{lipsum}
\usepackage{fancyhdr}       
\usepackage{graphicx}       
\graphicspath{{media/}}     

\usepackage{amsmath,amssymb,amsfonts}
\usepackage{booktabs}
\usepackage{algorithm}
\usepackage{algpseudocode}
\usepackage{setspace}
\newcommand{\overbar}[1]{\mkern 1.5mu\overline{\mkern-1.5mu#1\mkern-1.5mu}\mkern 1.5mu}
\usepackage{float,tabularx,booktabs,amsmath,mleftright}
\usepackage{lipsum}
\newcommand{\bftab}{\fontseries{b}\selectfont}

\usepackage{color}
 \usepackage{ulem}

\makeatletter
\def\@fnsymbol#1{\ensuremath{\ifcase#1\or \dagger\or *\or
   \mathsection\or \mathparagraph\or \|\or **\or \dagger\dagger
   \or \ddagger\ddagger \else\@ctrerr\fi}}
\makeatother
  
\title{SDGCCA: Supervised Deep Generalized Canonical Correlation Analysis for Multi-omics Integration}
\author{
  Sehwan Moon\thanks{Equal Contribution}\\
  School of Electrical Engineering and Computer Science \\
  Gwangju Institute of Science and Technology \\
  Gwangju\unskip,  South Korea\\
  \texttt{sehwanmoon@gm.gist.ac.kr} \\
  \And
  Jeongyoung Hwang\footnotemark[1]\\
  Artificial Intelligence Graduate School \\
  Gwangju Institute of Science and Technology \\
  Gwangju\unskip,  South Korea\\
  \texttt{wjddyd66@gm.gist.ac.kr} \\
  \AND
  Hyunju Lee\thanks{Corresponding author}\\
  School of Electrical Engineering and Computer Science, Artificial Intelligence Graduate School \\
  Gwangju Institute of Science and Technology \\
  Gwangju\unskip,  South Korea\\
  \texttt{hyunjulee@gist.ac.kr} \\
}

\begin{document}
\maketitle

\begin{abstract}
Integration of multi-omics data provides opportunities for revealing biological mechanisms related to certain phenotypes. We propose a novel method of multi-omics integration called supervised deep generalized canonical correlation analysis (SDGCCA) for modeling correlation structures between nonlinear multi-omics manifolds, aiming for improving classification of phenotypes and revealing biomarkers related to phenotypes.
SDGCCA addresses the limitations of other canonical correlation analysis (CCA)-based models (e.g., deep CCA, deep generalized CCA) by considering complex/nonlinear cross-data correlations and discriminating phenotype groups. Although there are a few methods for nonlinear CCA projections for discriminant purposes of phenotypes, they only consider two views. On the other hand, SDGCCA is the nonlinear multiview CCA projection method for discrimination.
When we applied SDGCCA to prediction of patients of Alzheimer's disease (AD) and discrimination of early- and late-stage cancers, it outperformed other CCA-based methods and other supervised methods. In addition, we demonstrate that SDGCCA can be used for feature selection to identify important multi-omics biomarkers. In the application on AD data, SDGCCA identified clusters of genes in multi-omics data, which are well known to be associated with AD.
\end{abstract}

\keywords{Canonical correlation analysis \and Deep neural networks \and Supervised learning \and Multi-omics \and Alzheimer’s disease}

\section{Introduction}
The advent of sequencing technology has facilitated the collection of genome-wide data for different molecular processes (e.g., gene expression, DNA methylation, microRNA [miRNA] expression), resulting in multi-omics data analysis  from the same set of individuals or biospecimens. Exploring molecular mechanisms using multi-omics data is expected to improve our current knowledge of diseases, which may lead to further improvements in disease diagnosis, prognosis,  and personalized treatment. While single-omics analysis can only capture a part of the biological complexity of a disease, integration of multi-omics data is required to provide a comprehensive overview of the underlying biological mechanisms.

Various methods such as unsupervised data integration models based on matrix factorization and correlation-based analysis, supervised data integration models based on network-based methods and multiple kernel learning, and Bayesian methods have been proposed for multi-omics data integration \cite{huang2017more}. For example, multi-omics factor analysis (MOFA)  \cite{argelaguet2018multi} is a Bayesian-based method for multi-omics integration by extracting the shared axes of variation between the different omics. Sparse generalized canonical correlation analysis (sGCCA) \cite{tenenhaus2014variable} is a generalization of regularized canonical correlation analysis with an L1-penalty model that selects co-expressed variables from omics datasets. Recently, researchers have been interested in multi-omics biomarkers that can explain or characterize a known phenotype. DIABLO (Data Integration Analysis for Biomarker discovery using Latent cOmponents) \cite{singh2019diablo} extends sGCCA to a supervised framework for identifying shared molecular patterns that can explain phenotypes across multi-omics; however, most of these methods are linear representations that cannot capture complex biological processes.

Canonical correlation analysis (CCA) \cite{hotelling1992relations} is a well-known multivariate model for capturing the associations between any two sets of data. CCA and its variations have been applied in several studies \cite{ tenenhaus2014variable,mandal2017faroc, jendoubi2019whitening, singh2019diablo} because of its advantages in biological interpretation. However, a drawback of CCA is that it can only consider the linear relationship of two modalities to maximally correlate them. Generalized canonical correlation analysis (GCCA) \cite{kettenring1971canonical} extends the CCA to the case of more than two modalities. To complement the GCCA, deep generalized canonical correlation analysis (DGCCA) \cite{benton2017deep} considers nonlinear relationship learning of more than two modalities. Also, supervised deep CCA \cite{liu2017supervised} and task-optimal CCA \cite{couture2019deep} have been proposed for supervised learning while considering nonlinear maximal correlation, but they can be only applied to two modalities.

In this study, we propose a supervised deep generalized canonical correlation analysis (SDGCCA), a nonlinear supervised learning model integrating with multiple modalities for discriminating phenotypic groups.
SDGCCA identifies the common and correlated information between multiple omics data, which  is important for discriminating phenotypic groups. SDGCCA is also based on a deep neural network (DNN), allowing the powerful capturing of the nonlinear part of the biological complexity. 
After training SDGCCA, we utilized Shapley additive explanation (SHAP) \cite{lundberg2017unified} to identify correlated biomarkers contributing to classification.

\section{Related Work}
In this section, we briefly review relevant previous studies.
Table \ref{tab:Table1} presents all the notions considered throughout the study .

\begin{table}[!ht]
	\begin{footnotesize}
		\vskip -0.15in
		\caption{The notations used in Eq. \ref{EQ:1}-\ref{EQ:12}}
		\vskip -0.2in
		\begin{center}
			\centering
			\setlength\tabcolsep{2pt}
			{\renewcommand{\arraystretch}{0.85}
				\begin{tabular}{lcl}
					\toprule
					\textsc{Notation} & \textsc{Dimension}      & \textsc{Description}\\ \midrule
					$n$               &-                        & Number of samples \\
					$m$               &-                        & Number of modalities \\
					$k$               &-                        & Dimensions of the shared representation \\
					$c$               &-                        & Number of label categories \\
					$d_i$             &-                        & Dimensions of the $i$-th modality. \\
					$\overbar{d_i}$   &-                        & Output dimensions of a deep neural network of the $i$-th modality. \\
					$f_i(\cdot)$      &-                        & Deep neural network of the $i$-th modality. \\ 
					$\theta_i$        &-                        & Parameters of $f_i(\cdot)$. \\ \midrule
					$X_i$             &$d_i \times n$           & $i$-th modality \\
					$V_i$             &$d_i \times k$           & Projection matrix for $X_i$ \\
					$U_i$             &$\overbar{d_i} \times k$ & Projection matrix for $f_i(X_i)$ \\
					$Y$               &$c \times n$             & Label \\
					$U_y$             &$c \times k$             & Projection matrix for $Y$ \\
					$U_y^{\dagger}$   &$c \times k$             & Pseudo inverse of $U_y$ \\
					$G$               &$k \times n$             & Shared representation \\\bottomrule
					\label{tab:Table1}
			\end{tabular}}
			\vskip -0.2in
		\end{center}
	\end{footnotesize}
\end{table}

\subsection{CCA}
CCA is one of the representative methods for dimension reduction that can consider the correlation between two modalities. 
It is trained to maximize the correlation between two mapped matrices using the projection matrices of each of the two modalities.
The objective function of CCA is as follows:
\begin{equation}
	(V_1^*, V_2^*) = \underset{V_1, V_2}{\text{argmax }} corr(V_1^\top X_1, V_2^\top X_2) = \underset{V_1, V_2}{\text{argmax }} \frac{V_1^\top \Sigma_{12} V_2}{\sqrt{V_1^\top \Sigma_{11} V_1 V_2^T \Sigma_{22} V_2}},\\
	\label{EQ:1}
\end{equation}
where $X_i$ denotes the $i$-th modality, $\Sigma_{11}$ and $\Sigma_{22}$ denote covariance matrices of $X_1$ and $X_2$, respectively, and $\Sigma_{12}$ denotes a cross-covariance matrix, $V_i$ denotes a projection matrix for the $i$-th modality, and $V_1^{*}$ and $V_2^{*}$ can be adopted to select the relevant features in both modalities.
Since the objective function above is invariant for scaling of $V_1$ and $V_2$, the final objective function is expressed as follows by adding the constraints of the unit variance.
\begin{equation}
	\begin{gathered}
		(V_1^*, V_2^*) = \underset{V_1, V_2}{\text{argmax }} V_1^\top \Sigma_{12} V_2, \\
		\text{s.t. } V_1^\top \Sigma_{11} V_1 = V_2^\top \Sigma_{22} V_2 = I
	\end{gathered}
	\label{EQ:2} 
\end{equation}
However, CCA has two limitations: (1) CCA is limited to mapping linear relationships and (2) CCA can only leverage  two modalities.

\subsection{DCCA}
A deep canonical correlation analysis (DCCA) \cite{andrew2013deep} is used to solve the limitations of CCA that extracts only the linear relationship.
In DCCA, to consider a nonlinear relationship, a DNN is applied to each modality.
DCCA is learned by maximizing the correlation of the DNN outputs of each modality.
The objective function of DCCA is as follows:
\begin{equation}
	({\theta_1}^*, {\theta_2}^*, {U_1}^*, {U_2}^*) = \underset{U_1, U_2}{\text{argmax }} corr(U_1^\top f_1(X_1), U_2^\top f_2(X_2)),
	\label{EQ:3} 
\end{equation}
where $f_i(.)$ is a DNN function for the $i$-th modality, $U_i$ indicates a projection matrix for $f_i(X_i)$, and $\theta_i$ is a parameter of $f_i(.)$.
$\theta_i(.)$ is trained via back-propagation that maximizes the objective function of DCCA.
However, because DCCA maximizes the correlation between the DNN outputs, unlike CCA, it cannot directly extract correlated features in both modalities. In addition, DCCA cannot be applied to more than two modalities.

\subsection{GCCA}
GCCA is used to extend the CCA to more than two modalities. The GCCA learns projection metrics that map each modality to a shared representation.
The objective function of GCCA is as follows:
\begin{equation}
	\begin{gathered}
		\underset{V_1, \ldots, V_m, G}{\text{minimize }} \sum_{i=1}^{m} \|G-V_i^{\top}X_i\|_F^2 ,\\
		\text{s.t. }GG^{\top} = I ,
		\label{EQ:4}
	\end{gathered}
\end{equation}
where $G$ denotes the shared representation and $V_i$ indicates a projection matrix for $X_i$. 
To solve the objective function of GCCA, an eigen decomposition of an $n \times n$ matrix is required, which increases quadratically with sample size and leads to memory constraints.
Also, unlike DCCA, nonlinear associations between modalities cannot be considered.

\subsection{DGCCA}
DGCCA is a model that addresses the two limitations of CCA by including both the advantages of GCCA and DCCA.
DGCCA learns projection metrics that map each output of DNN to a shared representation.
The objective function of DGCCA is as follows:
\begin{equation}
	\begin{gathered}
		\underset{U_1, \ldots, U_m, G}{\text{minimize }} \sum_{i=1}^{m} \|G-U_i^{\top} f_i(X_i)\|_F^2, \\
		\text{s.t. }GG^{\top} = I.
		\label{EQ:5}
	\end{gathered}
\end{equation}
$U_i$ and $G$ are trained to reduce the reconstruction error of GCCA, and to update $\theta_i$, gradients are back-propagated through the neural network. The gradient propagating to $f_i(X_i)$ is defined as $2U_i G - 2U_i U_i^\top f_i(X_i)$, and $\theta_i$ can be updated with  back-propagation to minimize the objective function of DGCCA.
	As $\theta_i$ is updated, the value of $f_i(X_i)$ is changed.
	Therefore, to solve the objective function of DGCCA, updating $U_i$ and $G$ and updating $\theta_i$ are alternately performed. 
DGCCA has the advantage of being able to obtain the nonlinear relationship of each modality. In addition, DGCCA can consider the correlation between more than two modalities.

\subsection{DIABLO}
DIABLO extends sGCCA, which is a GCCA  with L1-penalty. It is different from sGCCA as (1) the correlation between linear combinations of multi-omics data is changed to covariance; and (2) unlike sGCCA, which is an unsupervised method, it is a supervised framework capable of classification by maximizing the covariance between multiple omics datasets, including phenotype information. 
The objective function of DIABLO is as follows:
\begin{equation}
    \begin{gathered}
        \underset{V_1, \ldots, V_m, U_y}{\text{maximize }} \sum_{i,j=1; i \neq j}^{m} D_{i,j} \text{ } cov(V_i^\top X_i, V_j^\top X_j) + \sum_{l=1}^m D_{l,y} \text{ } cov(V_l^\top X_l, U_y^\top Y), \\
		\text{s.t. } \|V_i\|_2 = 1 \text{ and } \|V_i\|_1=\lambda_i ,
		\|U_y\|_2 = 1 \text{ and } \|U_y\|_1=\lambda_y,
		\label{EQ:6}
	\end{gathered}
\end{equation}
where $D=\{D_{ij}\} \in \mathbb{R}^{(m+1) \times (m+1)}$ is a design matrix that determines whether datasets should be connected.
However, DIABLO has a limitation—only assumes a linear relationship between the selected features to explain the phenotype.

\section{Methods} 
	\begin{figure}[ht!]
		\includegraphics[scale=1]{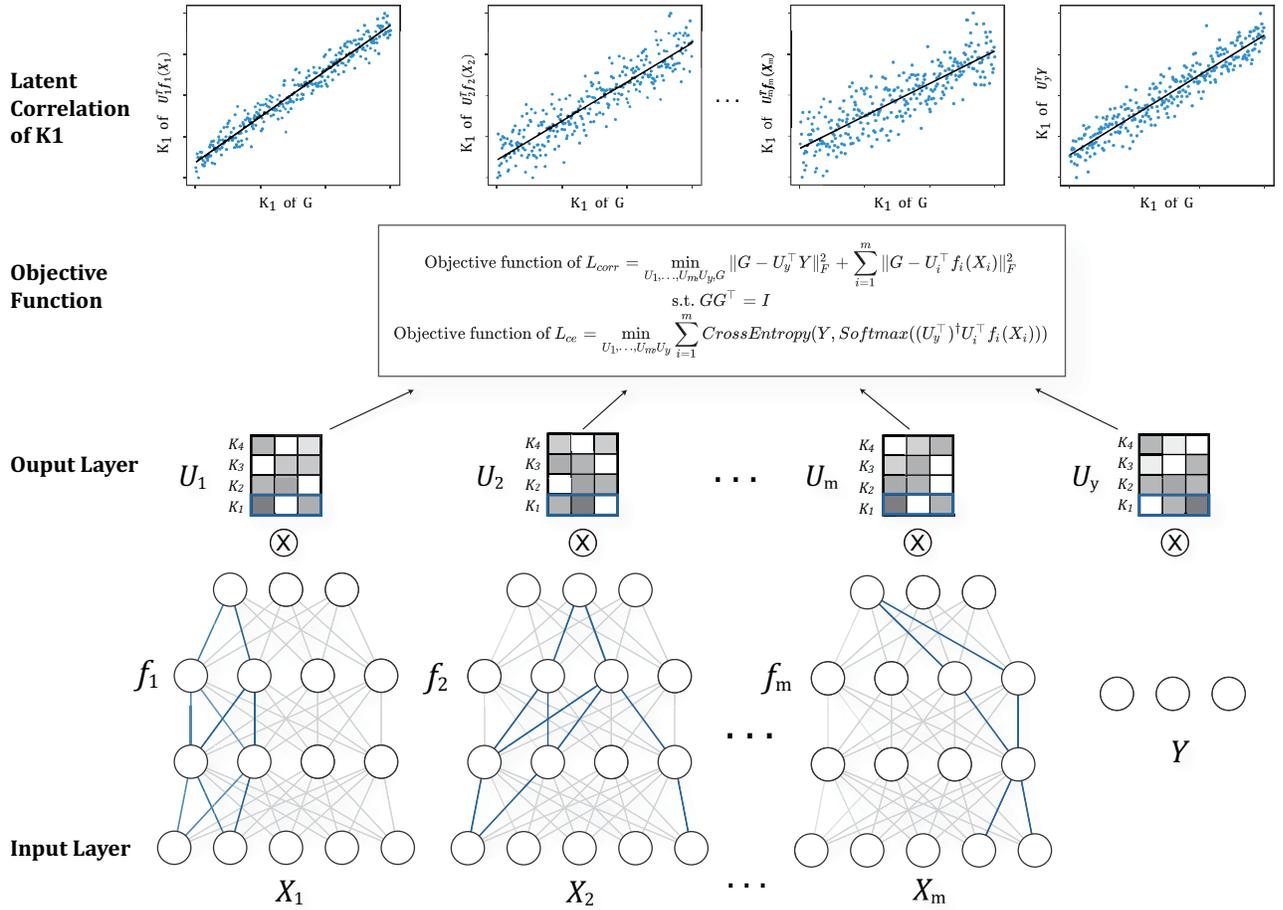}
		\vskip -2.3in
		\caption{\textbf{A schematic of SDGCCA.} $X_1,...,X_m$ are m modality, and $Y$ is the label information. Deep neural networks $f_1,...,f_m$ operate on $X_1,...,X_m$. The outputs of each modality and $Y$ are multiplied by each projection matrix ($U_1,...,U_m,U_y$). Two objective functions search for the optimal network $f_1,...,f_m$ and projection matrices, which provide both the highest correlation and lowest prediction error.} \label{fig1}
	\end{figure}
\subsection{The SDGCCA method}
The SDGCCA proposed in this study integrates ideas from DGCCA and DIABLO. SDGCCA incorporates the phenotypes of samples for supervised learning and  selects significant features based on CCA. It uses DNN to consider nonlinear interactions between multi-omics data including phenotype (Fig.~\ref{fig1}). SDGCCA made it possible to predict the phenotypes of samples by adding two elements to DGCCA.
First, the correlation of each modality and the correlation with the labels are considered. Thus, the shared presentation $G$ can be trained to obtain label information. The correlation loss function is defined as follows:
\begin{equation}
	\begin{gathered}
		L_{corr} = \|G-U_y^{\top}Y\|_F^2 + \sum_{i=1}^{m} \|G-U_i^{\top} f_i(X_i)\|_F^2 ,\\
		\text{s.t. }GG^{\top} = I ,
		\label{EQ:7}
	\end{gathered}
\end{equation}
where $U_y$ denotes a projection matrix for label $Y$.
Second, cross entropy \cite{de2005tutorial}, which is widely used in supervised models, is used to enable the propagation of label information directly to the DNN of each modality.
The projection matrix $U_y$ obtained from Eq. \ref{EQ:7} can map the label to the shared representation.
In addition, a projection matrix $U_i$ maps each modality to the shared presentation. 
Using the pseudo-inverse matrix of a projection matrix $U_y$, the label $Y$ can be approximated as follows: 
\begin{equation}
	\begin{gathered}
		G \approx U_i^{\top} f_i(X_i) \approx U_y^{\top}Y,\\
		Y \approx {(U_y^{\top})^{\dagger}} U_i^{\top} f_i(X_i),
		\label{EQ:8}
	\end{gathered}
\end{equation}
where $U_y^{\dagger}$ denotes the pseudo inverse of $U_y$. Then, let $\hat{Y_i}$=${(U_y^{\top})^{\dagger}} U_i^{\top} f_i(X_i)$. By applying a softmax function to $\hat{Y_i}$, the model is trained using cross entropy.
A classification loss can be defined as follows:
\begin{equation}
	\begin{gathered}
		L_{ce} = \sum_{i=1}^m CrossEntropy(Y, Softmax(\hat{Y_i})).
		\label{EQ:9}
	\end{gathered}
\end{equation}
The final label prediction of SDGCCA uses the soft voting of the label presentation ($\hat{Y_i}$) of each modality. The label prediction of SDGCCA is defined as follows:
\begin{equation}
	\hat{Y} = Softmax((\sum_{i=1}^m \hat{Y_i})/m),
	\label{EQ:10}
\end{equation}
where $m$ denotes the number of modalities.
The optimization of the proposed model consists of three main steps.
First, $U_i,...,U_m, U_y$ and $G$ are trained by the correlation loss function ($L_{corr}$). Here,	$G$ is obtained by solving an eigenvalue problem.
Let $C_{ii}=f_i(X_i) f_i(X_i)^\top, \text{s.t. }i=1, \ldots, m$, $C_{(m+1)(m+1)}=YY^\top$, $P_i=f_i(X_i)^\top C_{ii}^{-1} f_i(X_i) \in \mathbb{R}^{n \times n}$, $P_{m+1}=Y^\top C_{(m+1)(m+1)}^{-1} Y \in \mathbb{R}^{n \times n}$, and $M=\sum_{i=1}^{m+1} P_i$. 
Then, rows of $G \in \mathbb{R}^{n \times k}$ are orthonormal as top $k$ eigenvectors of $M$. 
If such $G$ is obtained, it can be easily obtained as $U_i = C_{ii}^{-1} f_i(X_i) G^\top$ and $U_y = C_{(m+1)(m+1)}^{-1} Y G^\top$.
Second, $\theta_i$ of $f_i(.)$ is trained using the $L_{corr}$.
It can be updated by selecting only the part related to $\theta_i$ in $L_{corr}$ and finding gradients to back-propagate to $f_i(X_i)$ as follows.
\begin{equation}
	\begin{gathered}
		\sum_{i=1}^{m} \|G-U_i^{\top} f_i(X_i)\|_F^2,\\
		=\sum_{i=1}^{m} \|G-Gf_i(X_i)^\top C_{ii}^{-1} f_i(X_i)\|_F^2,\\
		=\sum_{i=1}^{m} \|G(I_n-P_i))\|_F^2,\\
		=\sum_{i=1}^{m} \text{Tr}[G(I_n-P_i)G^\top],\\
		=\sum_{i=1}^{m} \text{Tr}(I_k) - \text{Tr}(GMG^\top),\\
		=Jk - \text{Tr}(GMG^\top).\\
	\end{gathered}
	\label{EQ:11}
\end{equation}
As above, $L_{corr}$ can be solved by maximizing $Tr(GMG^\top)$, and the derivative of $Tr(GMG^\top)$ with respect to $f_i(X_i)$ is demonstrated in DGCCA \cite{benton2017deep} as $2U_i G - 2U_i U_i^\top f_i(X_i)$.
Finally, after substituting the $U_i$ and $U_y$ obtained above into Eq. \ref{EQ:7}, $\theta_i$ is trained using $L_{ce}$.
A detailed algorithm for training SDGCCA using Eq. \ref{EQ:7}-\ref{EQ:11} is summarized in Algorithm 1.

	{\renewcommand{\arraystretch}{1.5}
		\begin{tabularx}{\textwidth}{ @{} X @{} }
			\toprule
			\normalsize{\textbf{Algorithm 1:} Training the proposed model }\\
			\midrule
			\hspace{0em}\textbf{Input:} 
			Training dataset $X=[X_1, X_2, \ldots, X_m]$, regularization rate $\alpha$, learning rate $\beta$, and max iterations T\\
			\hspace{0em}\textbf{Output:} Projection matrices $U_1, \ldots, U_m, U_y$, parameters $\theta_i$ of $f_i$ \\
			\hspace{1em}t $ = 1$ \\
			\hspace{1em}\textbf{while:} Validation loss does not converge or t $\leqq$ T\\
			\hspace{3em}\textbf{Step 1. Calculate $U_1, \ldots, U_m, U_y, G$} \\
			\hspace{4em}\begin{tabular}{ @{\hspace{\tabcolsep}} | l } 
				\hspace{0.5em} $L_{corr} = \|G-U_y^{\top}Y\|_F^2 + \sum_{i=1}^{m} \|G-U_i^{\top} f_i(X_i)\|_F^2$ \\
				\hspace{0.5em} $U_1, \ldots, U_m, U_y, G = \underset{U_1, \ldots, U_m, U_y, G}{\text{argmin }} L_{corr}$  
				\hspace{4em}\end{tabular} \\
			\hspace{3em}\textbf{Step 2. Training $\theta_i$ using $L_{corr}$} \\
			\hspace{4em}\begin{tabular}{ @{\hspace{\tabcolsep}} | l }
				\hspace{0.5em}$\nabla_{f_i(X_i)} L_{corr} \leftarrow{} U_iU_i^\top f_i(X_i)-U_iG$ \\
				\hspace{0.5em}$\theta_i \leftarrow{} (1-\alpha)\theta_i - \beta\nabla_{\theta_i} \nabla_{f_i(X_i)} L_{corr}$
				\hspace{4em}\end{tabular} \\
			\hspace{3em}\textbf{Step 3. Training $\theta_i$ using $L_{ce}$} \\
			\hspace{4em}\begin{tabular}{ @{\hspace{\tabcolsep}} | l }
				\hspace{0.5em}$\hat{Y}_i \leftarrow{} (U_y^{\top})^{\dagger} U_i^{\top} f_i(X_i)$\\
				\hspace{0.5em}$L_{ce} = \sum_{i=1}^m CrossEntropy(Y, Softmax(\hat{Y}_i))$\\
				\hspace{0.5em}$\theta_i \leftarrow{} (1-\alpha)\theta_i - \beta\nabla_{\theta_i} L_{ce}$
				\hspace{4em}\end{tabular} \\
			\hspace{3em} t $ \leftarrow{}$t $+ 1$ \\
			\hspace{1em}\textbf{end while}\\
			\bottomrule
	\end{tabularx}}

\subsection{Identification of multi-omics biomarkers}
SDGCCA is trained by maximizing the correlation of the DNN output, and a projection matrix can be used to select most correlated output from the DNN of each modality.
Because SDGCCA uses eigen decomposition to obtain a projection matrix as a CCA-based model, it can be observed that the correlation value of the first component ($U_i[:,1]$) is the largest among the values mapped to the shared representation through each modality.
Therefore, the DNN output with the highest correlative output among the DNN outputs corresponds to the maximum coefficient of the projection matrix (${\text{argmax }\lvert U_i[:,1] \rvert}$).
The most correlative output among each DNN is as follows:
\begin{equation}
	f_i(\cdot)[\text{argmax }\lvert U_i[:,1]\rvert, :]
	\label{EQ:12}
\end{equation}

However, unlike CCA and GCCA, the model is difficult to interpret due to the DNN. 
Thus, we used SHAP to select features related to the most correlative output among each DNN.
In \cite{lundberg2017unified}, SHAP calculates the feature importance using SHAP value that satisfies the desirable properties (local accuracy, missingness, and consistency) for each prediction. Specifically, we used Deep SHAP, which is tailored to DNN and effectively combines SHAP values calculated for smaller components of a DNN into SHAP values for the whole DNN.

\section{Results}
\subsection{Datasets}
We applied the proposed method to an AD classification task using multi-omics data. Three types of omics data (i.e. mRNA, DNA methylation, and microRNA (miRNA)) and clinical data were obtained from the ROSMAP cohort in the AMP-AD Knowledge Portal (\url{https://adknowledgeportal.synapse.org/}). We downloaded mRNA data that were normalized with quantile normalization to fragments per kilobase of transcript per million mapped read (FPKM) and removed potential batch effects using the Combat \cite{johnson2007adjusting}. The $\beta$-values of the downloaded DNA methylation data were measured using the Infinium HumanMethylation450 BeadChip and the missing $\beta$-values were imputed using a k-nearest neighbor algorithm.   We downloaded miRNA data that were normalized using variant stabilization normalization  and removed potential batch effects using the Combat. AD patients (n = 207) and normal controls (n = 169) with gene expression (GE), DNA methylation (ME), and miRNA expression (MI) profiles were included. 
Normalized FPKM values of the GE profiles were log2-transformed. For ME data, CpG sites located in promoter regions (TSS200 or TSS1500) were mapped to the corresponding gene, and the $\beta$-values of all overlapping genes were averaged. The MI profile was normalized using a variant stabilization normalization method, and batch effects were corrected using Combat \cite{johnson2007adjusting}. Finally, 18,164 GE features, 19,353 ME features, and 309 MI features were obtained.

To further measure the performance of the proposed method, we used kidney renal clear cell carcinoma (KIRC) data collected from The Cancer Genome Atlas (TCGA) for the early- and late-stage classification. The TCGA level-3 data on gene expression (Illumina mRNAseq), DNA methylation (Illumina HumanMethylation450 BeadArray), and miRNA expression (IlluminaHiSeq miRNAseq) were obtained. The methylation data used in this study were preprocessed according to \cite{ma2020diagnostic}. Finally, KIRC data are comprised of 313 samples (184 and 129 early- and late-stage samples, respectively) on 16,406 GE, 16,459 ME, and 342 MI features.

\subsection{Existing methods for performance comparison}
We compared the classification performance of the SDGCCA with the following ten existing methods. Here, we selected widely utilized machine learning or deep learning models and CCA-based multi-omics integration methods to relay how SDGCCA can contribute to the CCA framework. We selected (1) support vector machine (SVM), (2) extreme gradient boosting (XGB) \cite{chen2015xgboost}, (3) logistic regression (LR), (4) random forest (RF) as a method of machine learning, and (5) DNN as methods of deep learning. For CCA-based methods, (6) GCCA, (7) DGCCA, and (8) DIABLO \cite{singh2019diablo} were selected. 
Because GCCA and DGCCA are unsupervised learning models, SVM was used as an additional classification model.
In addition, the performance was compared with (9) Multi-Omics Graph cOnvolutional NETworks (MOGONET) \cite{wang2021mogonet} and (10) SMSPL \cite{9146338}, which are recently released multi-omics integration algorithms, although it is not a CCA-based model.

The performance of all combinations of GE, ME, and MI of ROSMAP, such as GE+ME, GE+MI, ME+MI, and GE+ME+MI was compared.
In addition, the performance of GE+ME+MI of KIRC was additionally compared.
We used accuracy (ACC), F1 score (F1), area under the receiver operating characteristic curve (AUC), and Matthews correlation coefficient (MCC) \cite{elith2006novel} as metrics for evaluating classification performance.
For all metrics, the mean and standard deviation for five-fold cross-validation (CV) were calculated.
Each CV used 60\% of the samples as a training set, 20\% as a validation set, and 20\% as a test set, and the hyperparameters of all models were selected based on the MCC of the validation set.

For SDGCCA, hyperparameters, including “Learning rate”  from the set \{$1\mathrm{e}{-4}$, $1\mathrm{e}{-5}$\}, “L2 regularization term on weights” from the set \{0, $1\mathrm{e}{-2}$, $1\mathrm{e}{-4}$\}, and “dimension of shared representation” from the set \{1, 2, $ \ldots$, 10\}, were selected using the validation set. Details about hyperparameters of all other models and five-fold cross validation are described in Supplementary Material.

SDGCCA is trained using correlation and classification losses. To see how each loss affects classification and feature selection, we performed ablation studies by measuring the performance of two additional models.
First, SDGCCA-$\text{G}_{corr}$ is a model excluding Step 2 of Algorithm 1 in the training process.
Second, SDGCCA-$\text{G}_{clf}$ is a model excluding Step 3 of Algorithm 1 in the training process.

\subsection{Evaluation of classification performances}
The results of the classification of AD patients and normal controls are summarized in Table \ref{tab:Table2}, \ref{tab:Table3}, \ref{tab:Table4}, and \ref{tab:Table5}. SDGCCA showed the best performance in 10 out of 16 cases, except for the performance of AUC in GE+ME, F1 in GE+MI, ACC, F1, MCC in ME+MI, and F1 in GE+ME+MI. In addition, for SDGCCA, the integration of all three omics data (GE+ME+MI) outperformed the integration of the two omics data. Interestingly, the integration of ME and MI showed different results from the combination of the other two omics data. For ME+MI, LR performed better than other machine learning models (SVM, XGB, and RF) and had the highest MCC values. In addition, SMSPL was the best performing model for ACC and FI measurements. If we consider that LR extracts the linear relationship between multi-omics data, and SMSPL is an LR-based model, the importance of nonlinearity in ME+MI is less than that in other combinations of omics data.

\begin{table*}[!hb]
	\begin{footnotesize}
		\caption{Performance comparison of AD classification using GE+ME in ROSMAP multi-omics data.}
		\begin{center}
			\begin{tabular}{lcccc}
				\toprule
				\textsc{Method}      & \textsc{ACC} & \textsc{F1} & \textsc{AUC} & \textsc{MCC}  \\ \midrule
				SVM          & 0.676 ± 0.044           & 0.711 ± 0.036          & 0.751 ± 0.055           & 0.346 ± 0.095           \\
				XGB          & 0.643 ± 0.063           & 0.686 ± 0.059          & 0.697 ± 0.053           & 0.275 ± 0.131           \\
				LR           & 0.674 ± 0.067           & 0.674 ± 0.072          & 0.750 ± 0.071           & 0.363 ± 0.133           \\
				RF           & 0.602 ± 0.058           & 0.687 ± 0.047          & 0.678 ± 0.059           & 0.179 ± 0.134           \\
				DNN          & 0.697 ± 0.037           & 0.695 ± 0.035          & 0.785 ± 0.038           & 0.412 ± 0.079           \\
				GCCA+SVM     & 0.665 ± 0.054           & 0.699 ± 0.046          & 0.710 ± 0.067           & 0.323 ± 0.111           \\
				DGCCA+SVM    & 0.609 ± 0.035           & 0.700 ± 0.021          & 0.673 ± 0.072           & 0.194 ± 0.080           \\
				DIABLO       & 0.633 ± 0.059           & 0.637 ± 0.062          & 0.702 ± 0.050           & 0.277 ± 0.120           \\
				MOGONET        & 0.670 ± 0.022           & 0.698 ± 0.050          & 0.698 ± 0.042           & 0.332 ± 0.034           \\ 
				SMSPL        & 0.683 ± 0.071           & 0.723 ± 0.056          & 0.751 ± 0.084           & 0.356 ± 0.155           \\ \midrule
				SDGCCA-$\text{G}_{corr}$ & 0.721 ± 0.050 & 0.724 ± 0.055 & \bftab  0.788 ± 0.043 & 0.453 ± 0.095 \\
				SDGCCA-$\text{G}_{clf}$ & 0.691 ± 0.034 & 0.693 ± 0.034 & 0.765 ± 0.046 & 0.396 ± 0.07 \\ \midrule
				SDGCCA       & \bftab 0.729 ± 0.035           & \bftab 0.728 ± 0.037      & 0.782 ± 0.019           & \bftab 0.474 ± 0.069          
				\\\bottomrule
				\multicolumn{5}{l}{\scriptsize The best performances are marked in bold.}
			\end{tabular}
			\label{tab:Table2}
		\end{center}
	\end{footnotesize}
\end{table*}

\begin{table*}[!hb]
	\begin{footnotesize}
		\begin{center}
			\caption{Performance comparison of AD classification using GE+MI in ROSMAP multi-omics data.}
			\begin{tabular}{lcccc}
				\toprule
				\textsc{Method}        & \textsc{ACC} & \textsc{F1} & \textsc{AUC} & \textsc{MCC} \\ \midrule
				SVM          & 0.679 ± 0.042           & 0.714 ± 0.036          & 0.755 ± 0.054           & 0.351 ± 0.089           \\
				XGB          & 0.647 ± 0.062           & 0.689 ± 0.057          & 0.704 ± 0.057           & 0.283 ± 0.130           \\
				LR           & 0.680 ± 0.069           & 0.681 ± 0.070          & 0.758 ± 0.070           & 0.375 ± 0.140           \\
				RF           & 0.602 ± 0.054           & 0.683 ± 0.046          & 0.678 ± 0.056           & 0.181 ± 0.126           \\
				DNN          & 0.689 ± 0.048           & 0.695 ± 0.049          & 0.765 ± 0.065           & 0.387 ± 0.095           \\
				GCCA+SVM     & 0.648 ± 0.044           & 0.700 ± 0.044          & 0.693 ± 0.060           & 0.288 ± 0.092           \\
				DGCCA+SVM    & 0.633 ± 0.071           & 0.714 ± 0.047          & 0.617 ± 0.095           & 0.244 ± 0.154           \\
				DIABLO       & 0.662 ± 0.060           & 0.672 ± 0.063          & 0.736 ± 0.066           & 0.330 ± 0.116           \\
				MOGONET        & 0.696 ± 0.055           & \bftab  0.722 ± 0.055          & 0.759 ± 0.040           & 0.387 ± 0.112           \\ 
				SMSPL        & 0.691 ± 0.079           & 0.719 ± 0.056          & 0.760 ± 0.062           & 0.378 ± 0.169           \\ \midrule
				SDGCCA-$\text{G}_{corr}$ & 0.697 ± 0.047 & 0.702 ± 0.053 & 0.757 ± 0.051 & 0.404 ± 0.091 \\
				SDGCCA-$\text{G}_{clf}$ & 0.667 ± 0.039 & 0.692 ± 0.030 & 0.739 ± 0.043 & 0.331 ± 0.082 \\ \midrule
				SDGCCA       & \bftab 0.699 ± 0.017           & 0.697 ± 0.015          & \bftab 0.796 ± 0.033           & \bftab 0.416 ± 0.035          
				\\\bottomrule
				\multicolumn{5}{l}{\scriptsize The best performances are marked in bold.}
			\end{tabular}
			\label{tab:Table3}
		\end{center}
	\end{footnotesize}
\end{table*}

\begin{table*}[!hb]
	\begin{footnotesize}
		\begin{center}
			\caption{Performance comparison of AD classification using ME+MI in ROSMAP multi-omics data.}
			\begin{tabular}{lcccc}
				\toprule
				\textsc{Method}        & \textsc{ACC} & \textsc{F1} & \textsc{AUC} & \textsc{MCC} \\ \midrule
				SVM          & 0.678 ± 0.040           & 0.713 ± 0.036          & 0.753 ± 0.052           & 0.349 ± 0.085           \\
				XGB          & 0.653 ± 0.059           & 0.697 ± 0.055          & 0.708 ± 0.054           & 0.296 ± 0.123           \\
				LR           & 0.683 ± 0.064           & 0.684 ± 0.065          & 0.758 ± 0.067           & \bftab  0.380 ± 0.130           \\
				RF           & 0.597 ± 0.051           & 0.682 ± 0.043          & 0.670 ± 0.058           & 0.169 ± 0.120           \\
				DNN          & 0.644 ± 0.033           & 0.637 ± 0.045          & 0.741 ± 0.031           & 0.305 ± 0.061           \\
				GCCA+SVM     & 0.631 ± 0.044           & 0.699 ± 0.037          & 0.672 ± 0.065           & 0.249 ± 0.096           \\
				DGCCA+SVM    & 0.561 ± 0.031           & 0.681 ± 0.027          & 0.548 ± 0.071           & 0.073 ± 0.079           \\
				DIABLO       & 0.686 ± 0.048           & 0.701 ± 0.051          & 0.755 ± 0.072           & 0.374 ± 0.095           \\
				MOGONET        & 0.668 ± 0.030           & 0.708 ± 0.040          & 0.708 ± 0.028           & 0.329 ± 0.048           \\ 
				SMSPL        & \bftab  0.686 ± 0.032           & \bftab 0.724 ± 0.025          & 0.747 ± 0.054           & 0.365 ± 0.068           \\ \midrule
				SDGCCA-$\text{G}_{corr}$ & 0.678 ± 0.050 & 0.679 ± 0.066 & \bftab 0.764 ± 0.052 & 0.369 ± 0.093 \\
				SDGCCA-$\text{G}_{clf}$ & 0.662 ± 0.012 & 0.681 ± 0.021 & 0.733 ± 0.029 & 0.325 ± 0.027 \\ \midrule
				SDGCCA       & 0.684 ± 0.046           & 0.693 ± 0.051          & \bftab  0.764 ± 0.039           & 0.372 ± 0.089          
				\\\bottomrule
				\multicolumn{5}{l}{\scriptsize The best performances are marked in bold.}
			\end{tabular}
			\label{tab:Table4}
		\end{center}
	\end{footnotesize}
\end{table*}

\begin{table*}[!th]
	\begin{footnotesize}
		\begin{center}
			\caption{Performance comparison of AD classification using GE+ME+MI in ROSMAP multi-omics data.}
			
			\begin{tabular}{lcccc}
				\toprule
				\textsc{Method}       & \textsc{ACC} & \textsc{F1} & \textsc{AUC} & \textsc{MCC}    \\
				\midrule
				SVM            & 0.679 ± 0.040& 0.714 ± 0.035 & 0.756 ± 0.050 & 0.352 ± 0.084 \\
				XGB            & 0.655 ± 0.060 & 0.698 ± 0.055 & 0.711 ± 0.055 & 0.299 ± 0.124 \\
				LR             & 0.683 ± 0.061 & 0.683 ± 0.063 & 0.759 ± 0.064 & 0.380 ± 0.124  \\
				RF             & 0.603 ± 0.050 & 0.684 ± 0.041 & 0.672 ± 0.055 & 0.181 ± 0.116 \\
				DNN            & 0.707 ± 0.039 & 0.701 ± 0.037 & 0.779 ± 0.043 & 0.437 ± 0.079 \\
				GCCA+SVM       & 0.628 ± 0.042 & 0.702 ± 0.033 & 0.669 ± 0.065 & 0.240 ± 0.094 \\
				DGCCA+SVM      & 0.569 ± 0.018 & 0.680 ± 0.037 & 0.615 ± 0.055 & 0.104 ± 0.034 \\
				DIABLO         & 0.673 ± 0.060 & 0.679 ± 0.064 & 0.739 ± 0.044 & 0.354 ± 0.117 \\
				MOGONET        & 0.684 ± 0.040           & 0.736 ± 0.012         & 0.692 ± 0.059           & 0.359 ± 0.086          \\ 
				SMSPL          & 0.699 ± 0.047 & 0.726 ± 0.027 & 0.777 ± 0.068 & 0.397 ± 0.110  \\ \midrule
				SDGCCA-$\text{G}_{corr}$ & \bftab 0.731 ± 0.035 & \bftab 0.742 ± 0.031 & 0.797 ± 0.034 & 0.469 ± 0.075 \\
				SDGCCA-$\text{G}_{clf}$ & 0.678 ± 0.047 & 0.682 ± 0.050 & 0.753 ± 0.061 & 0.367 ± 0.089 \\ \midrule
				SDGCCA         & \bftab 0.731 ± 0.050 & 0.729 ± 0.056 & \bftab 0.805 ± 0.043 & \bftab 0.479 ± 0.094 \\
				\bottomrule
				\multicolumn{5}{l}{\scriptsize The best performances are marked inbold.}
			\end{tabular}
			\label{tab:Table5}
		\end{center}
	\end{footnotesize}
\end{table*}

In all the experiments, SVM that uses the original input data performed better than GCCA+SVM and DGCCA+SVM.
In addition, SDGCCA performed better than GCCA+SVM and DGCCA+SVM, except for F1 in GE+MI.
This result indicates that there is a risk of losing information related to classification when dimension reduction is performed by only considering the correlation.
In most cases, the performance of SDGCCA-$\text{G}_{corr}$ was better than that of SDGCCA-$\text{G}_{clf}$, and the performances were improved when both the correlation and classification losses were combined.

The results of the classification of early-stage and late-stage of KIRC are shown in Table \ref{tab:Table6}. SDGCCA showed the best performance in two out of four cases, except for the performance of F1, and AUC.
F1 and AUC were the highest in LR-based SMSPL, and LR also had all higher performance than other machine learning models (SVM, XGB, and RF) and DNN.
Consistent with the results of ROSMAP, all performances of DGCCA+SVM were higher than those of GCCA+SVM, and all performances of SDGCCA-$\text{G}_{corr}$ were higher than those of SDGCCA-$\text{G}_{clf}$.

\begin{table*}[!th]
	\begin{footnotesize}
		\begin{center}
			\caption{Performance comparison of early- and late-stage classification using GE+ME+MI in KIRC multi-omics data.}
			
			\begin{tabular}{lcccc}
				\toprule
				\textsc{Method}       & \textsc{ACC} & \textsc{F1} & \textsc{AUC} & \textsc{MCC}    \\
				\midrule
				SVM            & 0.713 ± 0.040&	0.708 ± 0.039&	0.790 ± 0.035&	0.401 ± 0.082 \\
				XGB            & 0.693 ± 0.055&	0.688 ± 0.057&	0.778 ± 0.066	&0.362 ± 0.125 \\
				LR             & 0.738 ± 0.053&	0.738 ± 0.052	&0.784 ± 0.039&	0.480 ± 0.106  \\
				RF             & 0.687 ± 0.024&	0.661 ± 0.032&	0.770± 0.031&	0.340 ± 0.054 \\
				DNN            & 0.687 ± 0.023&	0.715 ± 0.025	&0.763 ± 0.054&	0.418 ± 0.041 \\
				GCCA+SVM       & 0.652 ± 0.057&	0.615 ± 0.073	&0.678 ± 0.086&	0.247 ± 0.159 \\
				DGCCA+SVM      & 0.665 ± 0.067&	0.642 ± 0.081	&0.684 ± 0.106&	0.287 ± 0.167 \\
				DIABLO         & 0.719 ± 0.052 & 0.760 ± 0.044 & 0.791 ± 0.030 & 0.425 ± 0.117 \\
				MOGONET        & 0.661 ± 0.095           & 0.728 ± 0.087         & 0.745 ± 0.061           & 0.327 ± 0.123          \\ 
				SMSPL          & 0.710 ± 0.069 &\bftab 0.763 ± 0.052 & \bftab0.808 ± 0.067 & 0.394 ± 0.151  \\ \midrule
				SDGCCA-$\text{G}_{corr}$ & 0.741 ± 0.063&	0.742 ± 0.062	&0.800 ± 0.058&	0.479 ± 0.129 \\
				SDGCCA-$\text{G}_{clf}$ & 0.735 ± 0.060&	0.734 ± 0.057&	0.794 ± 0.061	&0.472 ± 0.122\\ \midrule
				SDGCCA         & \bftab0.745 ± 0.035&	0.745 ± 0.034&	0.793 ± 0.084&	\bftab 0.484 ± 0.069 \\
				\bottomrule
				\multicolumn{5}{l}{\scriptsize The best performances are marked in bold.}
			\end{tabular}
			\label{tab:Table6}
		\end{center}
	\end{footnotesize}
\end{table*}

To statically estimate the performance of our model against other models, we performed a paired $t$-test using five-fold cross-validation classification results in MCC values for GE+ME+MI of ROSMAP and KIRC (Table 7). We found that SDGCCA statistically outperformed its competing methods in 15 of the 20 cases ($p$-value $<$ 0.05).

\begin{table*}[!th]
	\begin{footnotesize}
		\begin{center}
			\caption{Statistical significances of performance improvements of SDGCCA against other methods.}
			
			\begin{tabular}{lcc}
				\toprule
				\textsc{Methods} & \textsc{ROSMAP} & \textsc{KIRC} \\ \midrule
				SVM & 6.57E-02 & \textbf{7.22E-04} \\
				XGB & \textbf{2.82E-02} & \textbf{2.28E-02} \\
				LR & \textbf{1.68E-03} & 4.40E-01 \\
				RF & 5.65E-02 & \textbf{1.09E-02} \\
				DNN & \textbf{8.39E-03} & \textbf{3.28E-02} \\
				GCCA+SVM & 9.26E-02 & \textbf{4.17E-03} \\
				DGCCA+SVM & \textbf{1.72E-02} & \textbf{1.35E-02} \\
				DIABLO & \textbf{1.67E-03} & \textbf{2.36E-02} \\
				MORONET & \textbf{1.59E-02} & \textbf{3.51E-02} \\
				SMSPL & \textbf{1.59E-02} & 5.24E-02\\
				\bottomrule
				\multicolumn{1}{l}{\scriptsize Values with a $p$-value $<$ 0.05 are marked in bold.}
			\end{tabular}
			\label{tab:Table7}
		\end{center}
	\end{footnotesize}
\end{table*}

\begin{figure}[hp!]
	\includegraphics[width=\textwidth]{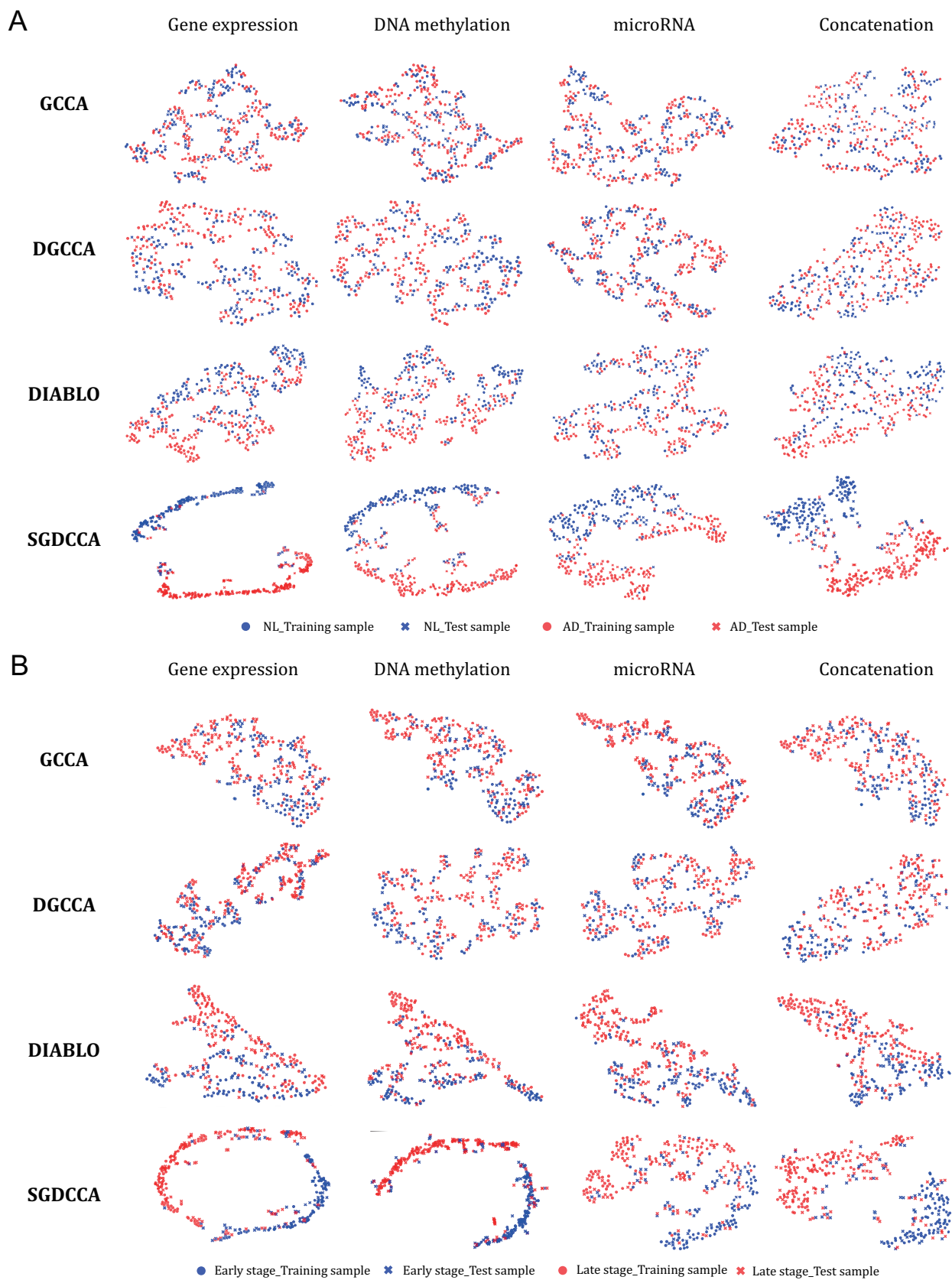}
	\caption{ \textbf{$t$-SNE plots for CCA-based methods including GCCA, DGCCA, DIABLO, and SGDCCA.} (A) ROSMAP data and (B) TCGA KIRC data. Each method was used to compute projections for the gene expression, DNA methylation, and miRNA data. The circle and cross symbols represent training and test samples, respectively. Samples are colored according to labels.
	} \label{fig2}
\end{figure}

We projected each omics data and concatenated multi-omics data on a low-dimensional space throughout dimension reduction by t-distributed Stochastic Neighbor Embedding (t-SNE). Figure~\ref{fig2} visualizes the projections of multi-omics data for each method. As expected, the supervised learning-based methods more clearly separated classes. Among the supervised learning-based models, nonlinear SGDCCA classified classes more clearly than linear DIABLO.

To further demonstrate the effects of the hyperparameter k (dimension of shared representation) on the SDGCCA, we trained SDGCCA under a wide range of k using the ROSMAP data. Figure S2 shows the embedding performance, correlation sum, and classification performance of SDGCCA when k varies from 1 to 10. we observed that the hyperparameter k did not influence the embedding performance and classification performance of SDGCCA as the performance fluctuated with the change of K. However, we observed that the correlation sum peaked at K=7 and decreased thereafter. This experiment described in detail in Supplementary material.

\subsection{Classification performance of the identified biomarkers}
We compared the feature selection performance of the CCA-based method to demonstrate that the set of relevant features of DNN output with high correlation between each modality is effective in classification.
For each CV, SHAP selected 300 out of 18,164 features for GE, 300 out of 19,353 features for ME, and 30 out of 309 features for MI using only training data of ROSMAP dataset.
Here, correlated features were selected using only training data.
Performance was evaluated using LR, which was the best performance among the machine learning models in the GE+ME+MI experiments.
To  confirm whether the features selected by SDGCCA are important features in the classification, LRs with these features were compared with randomly selected features and features selected by CCA-based models, GCCA and DGCCA . The comparisons were repeated 100 times.

\begin{table*}[!ht]
	\begin{footnotesize}
		\begin{center}
			\caption{ROSMAP classification performance comparison of important features selected by CCA-based methods.}
			\label{table1}
			
			\begin{tabular}{lcccc}
				\toprule
				\textsc{Feature Set}       & \textsc{ACC} & \textsc{F1} & \textsc{AUC} & \textsc{MCC}    \\
				\midrule
				All Features    & 0.683 ± 0.061 & 0.683 ± 0.063 & \bftab 0.759 ± 0.064 & 0.380 ± 0.124 \\
				Random Features & 0.661 ± 0.043 & 0.674 ± 0.047 & 0.727 ± 0.045 & 0.328 ± 0.086 \\
				GCCA            & 0.630 ± 0.043 & 0.638 ± 0.036 & 0.693 ± 0.045 & 0.269 ± 0.089 \\
				DGCCA           & 0.669 ± 0.040 & 0.678 ± 0.048 & 0.739 ± 0.037 & 0.345 ± 0.076 \\ \midrule
				SDGCCA-$\text{G}_{corr}$ & 0.646 ± 0.045 & 0.661 ± 0.048 & 0.716 ± 0.053 & 0.294 ± 0.092 \\
				SDGCCA-$\text{G}_{clf}$ & 0.650 ± 0.021 & 0.650 ± 0.020 & 0.739 ± 0.040 & 0.315 ± 0.048 \\ \midrule
				SDGCCA          & \bftab 0.689 ± 0.045 & \bftab 0.698 ± 0.042 & 0.755 ± 0.043 & \bftab 0.386 ± 0.095 \\ \bottomrule
				\multicolumn{5}{l}{\scriptsize The best performances are marked in bold.}
			\end{tabular}
			\label{tab:Table8}
		\end{center}
	\end{footnotesize}
\end{table*}

Table \ref{tab:Table8} presents the classification performance of important features selected by all the competing methods. 
All features and feature sets obtained from DGCCA and SDGCCA performed better than the randomly selected features, while the other feature sets did not. A feature set from SDGCCA showed better performance than using all features except AUC. Thus, it can be observed that SDGCCA can identify important features in AD classification using multi-omics data.

SDGCCA-$\text{G}_{corr}$ and SDGCCA-$\text{G}_{clf}$ performed lower than when using randomly selected features. 
Regarding SDGCCA-$\text{G}_{corr}$, the gradient associated with the correlation is not propagated to the weight and bias of the DNN of each modality, indicating that the ability to select the correlative features between multi-omics is worse.
When we calculated the average of correlations between the first components of shared presentation of each modality in the training set, SDGCCA-$\text{G}_{corr}$ had a correlation coefficient of 0.462, which is much lower than the correlation coefficient of 0.954 from SDGCCA-$\text{G}_{clf}$, and the correlation coefficient of 0.956 from SDGCCA.
Regarding SDGCCA-$\text{G}_{clf}$, the correlation value is slightly lower than that of SDGCCA, but shows lower performance.
Accordingly, $L_{corr}$ only cannot propagate sufficient information about the label to the DNN of each modality, and it is important to use $L_{clf}$ together.

\subsection{Pathway analysis using the SHAP values}
\begin{figure}[ht!]
	\includegraphics[width=\textwidth]{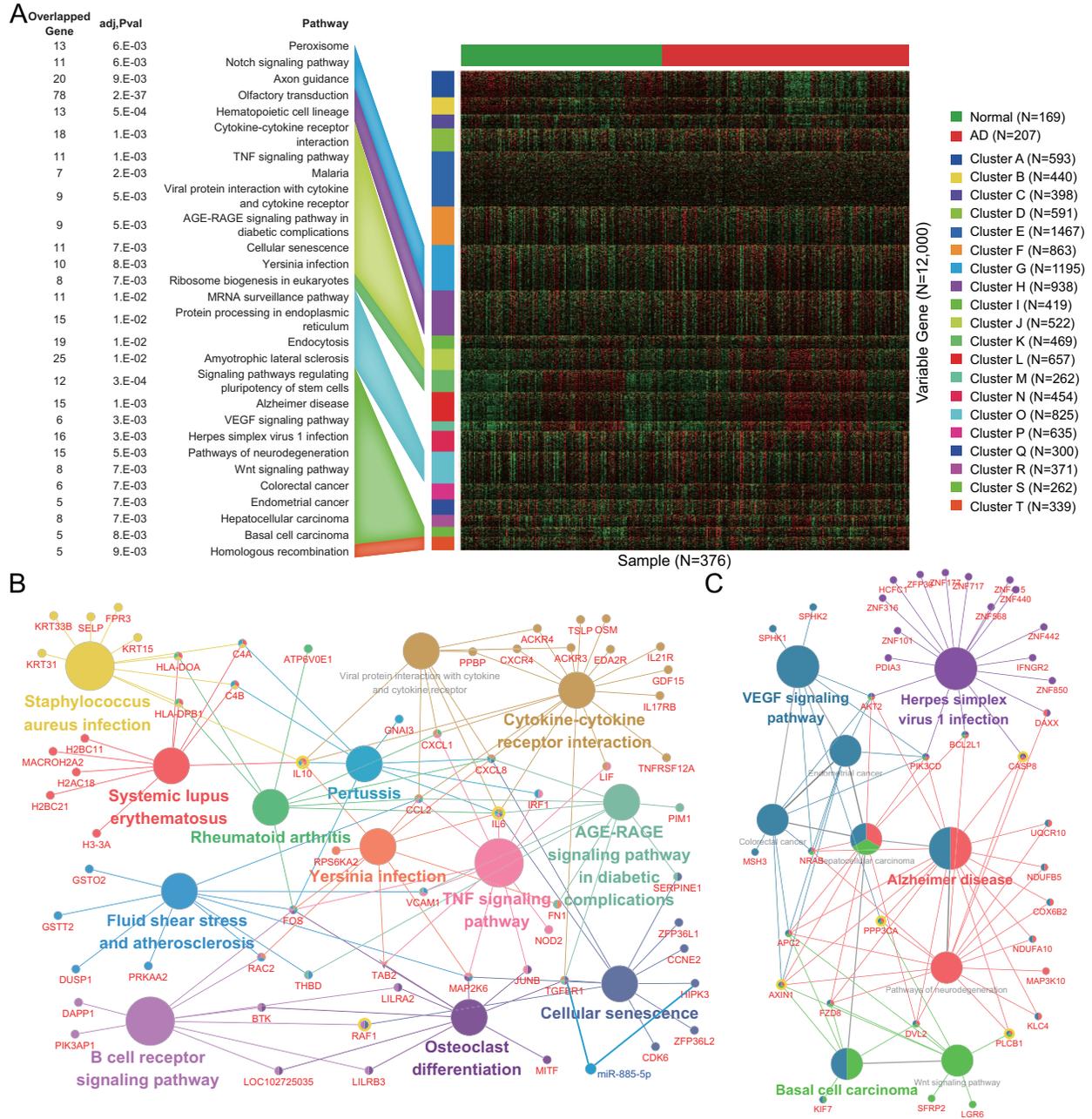}
	\caption{ \textbf{The results of pathway analysis.} (A) Clustered heatmap of SHAP value with red color denoting increase and green color denoting decrease. The Kyoto Encyclopedia of Genes and Genomes pathways enriched in each of the 20 clusters are represented with a heatmap. (B) The pathway network of cluster J. (C) The pathway network of cluster S. Yellow circles denote well-known Alzheimer's disease-related genes.
	} \label{fig3}
\end{figure}

To further illustrate the applicability of the proposed method, we performed pathway analysis. For the pathway analysis, all ROSMAP samples were used for training the SDGCCA, where hyperparameters having the highest MCC values for five folds on average in the five-fold cross-validation were used.
We clustered features with similar patterns using all the samples and features with variable SHAP values (Fig.~\ref{fig3} (A)). Pathway enrichment analysis was performed based on the Kyoto encyclopedia of genes and  genomes (KEGG) database \cite{kanehisa2000kegg} with the GE and ME features of each cluster. Fig.~\ref{fig3}A illustrates enriched KEGG pathways with adjusted p-values of less than 0.05. Cluster H was significantly enriched in the KEGG pathway related to olfactory transduction (adjusted P-values $=$ 2.E-37). Previous studies \cite{zou2016olfactory} have revealed that AD is closely related to olfactory dysfunction in AD. We analyzed clusters J and S in detail using ClueGO \cite{bindea2009cluego} Cytoscape plugins to show the relationship (Fig.~\ref{fig3} (B) and (C)). In cluster J, we identified that the two genes HIPK3 and TGFBR1 related to cellular senescence, and miR-885-5p targeting them were clustered. Cellular senescence is widely known to be associated with AD \cite{boccardi2015cellular,masaldan2019cellular,reddy2017micrornas}. In addition, IL6, IL10 and RAF1 genes in cluster J network are also well-known AD closely related genes \cite{arosio2004interleukin, mei2006distribution}. Cluster S was significantly enriched in the KEGG pathway related to AD (adjusted P-values $=$ 1.E-03) and neurodegeneration (adjusted P-values $=$ 5.E-03). CASP8 and PLCB1 are  known as AD biomarkers \cite{rehker2017caspase, shimohama1995signal}. AXIN1 and PPP3CA are also known as AD-related genes \cite{lloret2011amyloid, whelan2019multiplex}.

\section{Conclusion}
In this study, we proposed a CCA-based SDGCCA, an integration method of multi-omics data for the classification and identification of significant multi-omics biomarkers.
SDGCCA was trained to consider the nonlinear/complex interaction between multi-omics using the loss of DGCCA that maximizes the correlation of each DNN output. 
In addition, because the label can be predicted using a projection matrix, it is possible to train the model to propagate label information to each DNN using cross entropy. SDGCCA performed better in the AD classification task using gene expression, DNA methylation, and miRNA expression than the other machine learning models, DNN, DIABLO, MOGONET, and SMSPL.
We showed that SDGCCA can select an important feature set related to a phenotype by comparing it with other feature selection models.
Using SHAP values, we performed clustering of features in multi-omics data, and showed that it is applicable to AD-related biomarker discovery using pathway analysis.
In conclusion, SDGCCA is a multi-omics integration algorithm with high classification performances and has the ability to select a set of mutually contributing features from different multi-omics datasets.

\section*{Software}
Source codes of SDGCCA are available at https://github.com/DMCB-GIST/SDGCCA.

\section*{Acknowledgments}
This work was supported by supported by the Bio \& Medical Technology Development Program of NRF funded by the Korean government (MSIT) (NRF-2018M3C7A1054935) and Institute of Information \& communications Technology Planning \& Evaluation (IITP) grant funded by the Korea government (MSIT) (No.2019-0-01842, Artificial Intelligence Graduate School Program (GIST)).

\bibliographystyle{unsrt}  
\bibliography{main}  

\pagebreak
\begin{center}
\textbf{\Large Supplemental Materials\\ SDGCCA: Supervised Deep Generalized Canonical Correlation
Analysis for Multi-omics Integration}
\end{center}
\setcounter{equation}{0}
\setcounter{figure}{0}
\setcounter{table}{0}
\setcounter{page}{1}
\setcounter{section}{0}

\makeatletter
\renewcommand{\thefigure}{S\arabic{figure}}

\thispagestyle{empty}

\section{Hyperparameters tuning}
\label{sub:sec1}
\begin{figure}[ht!]
\includegraphics[width=\textwidth]{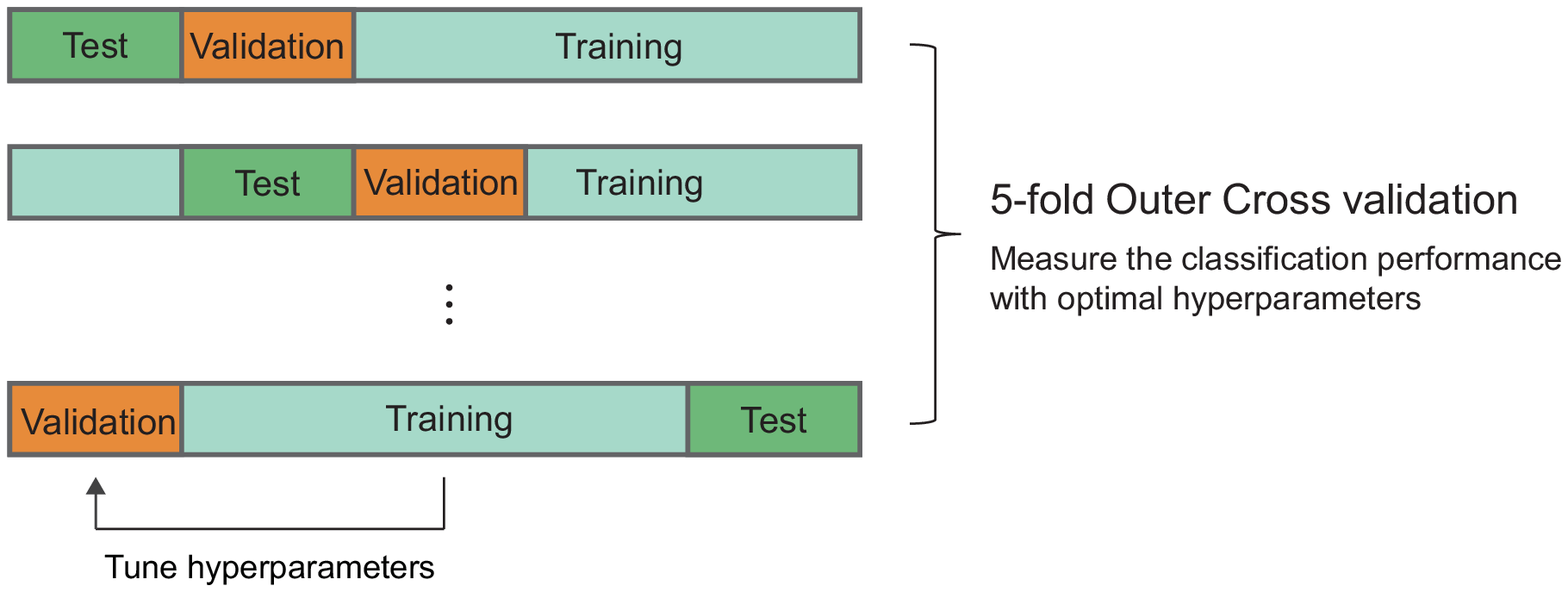}
\caption{ \textbf{Schematic of five-fold cross-validation (CV)} The data are divided into five equal sets. At each split, the model is trained on three sets and validated on one set. The last set measures the performance of the model. 
}\label{fig4}
\end{figure}
We compared the proposed model with other models for classifying normal control and Alzheimer's disease in ROSMAP and classifying early- and late-stage samples in KIRC. 
Figure~\ref{fig4} describes five-fold cross validation used in this study.
The values of hyperparameters of all models were selected based on Matthews correlation coefficient (MCC) performance of the validation set, and sets of hyperparameter values are as follows:
\begin{itemize}
    \item Support vector machine (SVM): The value of a parameter ‘Kernel' is chosen from \{poly, rbf, linear\} and that of ‘L2 regularization term on weights' is from five equally divided values from \{$1\mathrm{e}{-3}$$\sim$$1\mathrm{e}{+3}$\}.
    
    \item Extreme gradient boosting (XGB) (\cite{chen2015xgboost}): The value of a parameter  ‘Maximum depth of a tree' is chosen from \{3, 5\}, that of ‘Subsample ratio of the training instances' are from \{0.6, 0.8\}, that of ‘Minimum sum of instance weight needed in a child' are from \{0.6, 0.8\}, and that of ‘L1 regularization term on weights' are from \{0.1, 0.5\}. 
    
    \item Logistic regression (LR): The value of a parameter ‘Type of regularization' is chosen from \{L1, L2\}, and that of ‘Regularization term on weights' is from  ten equally divided values from \{$1\mathrm{e}{-5}$$\sim$$1\mathrm{e}{+5}$\}.
    
    \item Random forest (RF): The value of a parameter ‘Function to measure the quality of a split' is chosen from \{gini impurity, information gain\}, that of ‘Minimum number of samples required to be at a leaf node' is from \{0.1, 0.5\}, that of ‘Maximum depth of the tree' is from \{3, 5, 7, 9\}, and that of ‘Minimum number of samples required to split an internal node' is from \{0.5, 0.7, 0.9\}.
    
    \item Deep neural network (DNN): The value of a parameter ‘Learning rate' is chosen  from \{$1\mathrm{e}{-4}$, $1\mathrm{e}{-5}$\}, and that of ‘L2 regularization term on weights' is from a set \{0, $1\mathrm{e}{-2}$, $1\mathrm{e}{-4}$\}.
    
    \item Generalized canonical correlation analysis (GCCA) (\cite{kettenring1971canonical}): The value of a parameter ‘Dimension of shared representation' is chosen  from the set \{1, 2, $ \ldots$, 10\}. Since GCCA is an unsupervised model, SVM is used as a classifier. The hyperparameters of  SVM are the same as above.
    
    \item Deep generalized canonical correlation analysis (DGCCA) (\cite{benton2017deep}): The value of a parameter  ‘Learning rate'  is chosen from \{$1\mathrm{e}{-4}$, $1\mathrm{e}{-5}$\}, that of ‘L2 regularization term on weights' is from the set \{0, $1\mathrm{e}{-2}$, $1\mathrm{e}{-4}$\}, and that of ‘Dimension of shared representation' is from \{1, 2, $ \ldots$, 10\}. Since DGCCA is an unsupervised model, SVM is used as a classifier. The hyperparameters of  SVM are the same as above. 
    
    \item Data Integration Analysis for Biomarker discovery using Latent cOmponents (DIABLO) (\cite{singh2019diablo}): The value of a parameter ‘Number of components‘  is chosen  from \{1,2,...,10\} based on MCC performance of the validation set. The value of a parameter ‘Number of features for each dataset' is chosen  from \{5,15,25,35,45\} based on centroid distance performance from the training set. 
    
    \item SMSPL (\cite{9146338}): The value of a parameter ‘Parameter for adjusting influence from other modalities’  is chosen  from  \{0.66,0.1,0.01\},  that of ‘Age parameter’ is from  \{(0.66, 0.66), (0.1, 0.1), (0.01, 0.01)\}, that of ‘Size of increasing the age parameter with each iteration’ is from  \{0.01, 0.02, 0.04, 0.08\}, and that of ‘Size to increase the selected sample for each iteration’ is from \{2, 4\}.
    
    \item Multi-Omics Graph cOnvolutional NETworks (MOGONET) (\cite{wang2021mogonet}): The value  of a parameter ‘Threshold of affinity values’  is chosen from \{2, 4, 6, 8, 10\}, that of ‘Learning rate for pretraining’ is from \{$5\mathrm{e}{-3}$, $5\mathrm{e}{-4}$\}, that of ‘Learning rate for graph convolutional network’ is from \{$5\mathrm{e}{-3}$, $5\mathrm{e}{-4}$\}, that of ‘Learning rate for classifier’ is from  \{$5\mathrm{e}{-3}$, $5\mathrm{e}{-4}$\}, and that of ‘Number of significant features for each omics data type’ was fixed as 200.
    
    \item Supervised Deep Generalized Canonical Correlation Analysis (SDGCCA): The value of a parameter ‘Learning rate'  is chosen from the set \{$1\mathrm{e}{-4}$, $1\mathrm{e}{-5}$\}, that of ‘L2 regularization term on weights' is from the set \{0, $1\mathrm{e}{-2}$, $1\mathrm{e}{-4}$\}, and that of ‘Dimension of shared representation' is from the set \{1, 2, $ \ldots$, 10\}.
   
 \item SDGCCA - $G_{corr}$: It has the same hyperparameter sets of SDGCCA.
    
\item SDGCCA - $G_{clf}$: It has the same hyperparameter sets of SDGCCA.
\end{itemize}

\section{Performance of SGCCA under various values of hyperparameter k.}

\begin{figure}[ht!]
\includegraphics[width=\textwidth]{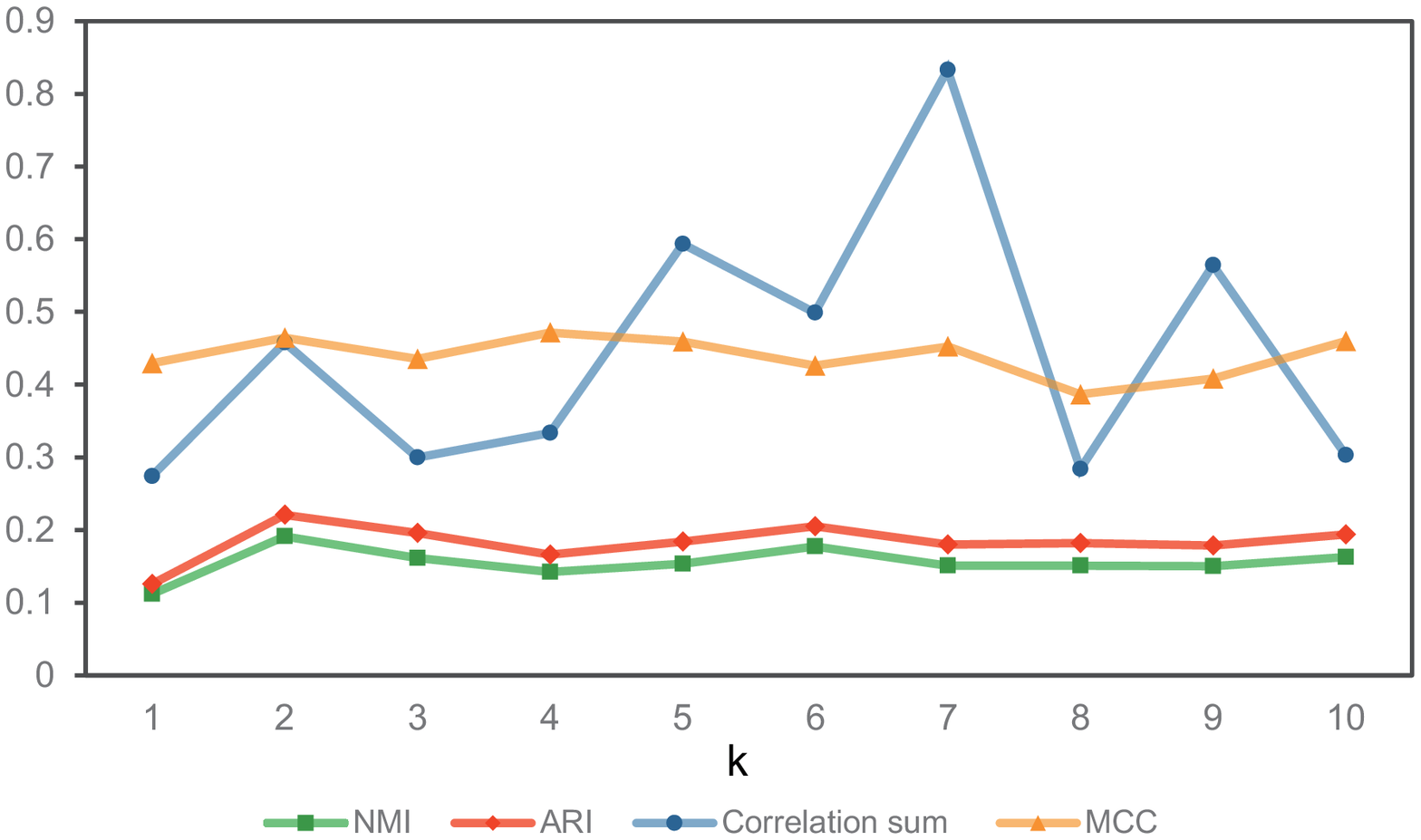}
\caption{ \textbf{Performance of SGCCA under various values of hyperparameter $k$ (dimension of shared representation) using ROSMAP dataset.} 
} \label{fig5}
\end{figure}

We measured the change of performances of SDGCCA with respect to a hyperparameter $k$ (dimension of shared representation).
Using the trained model, a total of four metrics were measured: normalized mutual information (NMI), adjusted rand index (ARI), correlation sum, and MCC.
NMI and ARI were measured using K-means clustering, where input is concatenation of the output of DNN mapped to the shared representation ($U_i^{\top} f_i(X_i) \in \mathbb{R}^{k \times n}$).
For the correlation sum, the correlation between the rows of each $U_i^{\top} f_i(X_i)$ was obtained, and the average of the three correlations (i.e. correlation of GE and ME, correlation of GE and MI, and correlation of ME and MI) was calculated. 
After that, the $k$ correlation values were summed.
The correlation sum is calculated as follows:
\begin{equation}
\begin{gathered}
\sum_{i=1}^{k} (\sum_{j,l=1;j \neq l}^{3} corr(U_j^{\top} f_j(X_j)[i, :], U_l^{\top} f_l(X_l)[i, :]))/3.
\label{EQ:4}
\end{gathered}
\end{equation}
MCC was measured by softvoting the prediction of each modality. 

Figure~\ref{fig5} shows the NMI, ARI, correlation sum, and MCC of SDGCCA when $k$ varies from 1 to 10 using the ROMAP dataset.
We observed that the hyperparameter $k$ did not influence the classification performance and embedding performance of SGDCCA, while the correlation sum peaked at $k$=7 and decreased thereafter.


\end{document}